\documentclass[sigconf]{acmart}
\setcopyright{none} 
\renewcommand\footnotetextcopyrightpermission[1]{}

\def\codex{Codex}
\def\hackerrank{HackerRank}

\usepackage{graphicx}
\usepackage{subcaption}
\usepackage{csquotes}
\usepackage{booktabs}
\usepackage{array}
\usepackage{hhline}

\newcommand\topspace{\rule{0pt}{2.0ex}}            
    
\newcommand\bottomspace{\rule[-0.9ex]{0pt}{0pt}}

\usepackage{listings}
\usepackage{xcolor, tcolorbox}
\definecolor{code}{rgb}{0.5,0.5,0.5}
\definecolor{codegreen}{rgb}{0,0.6,0}
\definecolor{codepurple}{rgb}{0.58,0,0.82}
\definecolor{backcolour}{rgb}{0.98,0.98,0.98}
\definecolor{bordercolour}{rgb}{0.95,0.95,0.95}
\definecolor{white}{rgb}{1,1,1}
\definecolor{gridvar}{rgb}{0.3,0.3,0.3}
\lstdefinestyle{mystyle}{
	backgroundcolor=\color{backcolour},   
	commentstyle=\color{codegreen},
	keywordstyle=\color{magenta},
	numberstyle=\tiny\color{codegray},
	stringstyle=\color{codepurple},
	rulecolor=\color{bordercolour},
	basicstyle=\ttfamily\footnotesize,
	breakatwhitespace=true,         
	breaklines=true,                 
	captionpos=b,                    
	keepspaces=true,                 
	numbers=left,               
	numbersep=5pt,                  
	showspaces=false,                
	showstringspaces=false,
	showtabs=false,
	framextopmargin=2pt,
	framexbottommargin=0pt,
	framexleftmargin=0pt,
	framexrightmargin=0pt,
}

\definecolor{my-blue}{cmyk}{0, 0, 0, 0.07, 1.00}
\newcommand{\resultbox}[1]{
	\begin{tcolorbox}[colback=my-blue,colframe=gray!30!white,top=3pt, left=3pt,right=3pt, bottom=3pt]
		#1
	\end{tcolorbox}
}

\begin{document}

\title{Codex Hacks HackerRank: Memorization Issues and a Framework for Code Synthesis Evaluation}

\author{Anjan Karmakar}
\affiliation{%
  \institution{Free University of Bozen-Bolzano}
  \city{Bolzano}
  \country{Italy}
  }
\email{akarmakar@unibz.it}

\author{Julian Aron Prenner}
\affiliation{%
  \institution{Free University of Bozen-Bolzano}
    \city{Bolzano}
  \country{Italy}
  }
\email{julianaron.prenner@unibz.it }

\author{Marco D'Ambros}
\affiliation{%
  \institution{CodeLounge, Software Institute}
    \city{Lugano}
  \country{Switzerland}
  }
\email{marco.dambros@usi.ch}

\author{Romain Robbes}
\affiliation{%
  \institution{Free University of Bozen-Bolzano}
    \city{Bolzano}
  \country{Italy}
  }
\email{rrobbes@unibz.it }

\begin{abstract}

The \codex{} model has demonstrated extraordinary competence in synthesizing code from natural language problem descriptions \citep{chen2021evaluating}. However, in order to reveal unknown failure modes and hidden biases, such large-scale models must be systematically subjected to multiple and diverse evaluation studies. 

In this work, we evaluate the code synthesis capabilities of the \codex{} model based on a set of 115 Python problem statements from a popular competitive programming portal: \hackerrank{}. Our evaluation shows that Codex is indeed \textit{proficient} in Python---solving 96\% of the problems in a zero-shot setting, and 100\% of the problems in a few-shot setting. However, \codex{} exhibits clear signs of generating memorized code based on our evaluation. This is alarming, especially since the adoption and use of such models could directly impact how code is written and produced in the foreseeable future. With this in mind, we further discuss and highlight some of the prominent risks associated with large-scale models of source code. Finally, we propose a framework for code-synthesis evaluation using variations of problem statements based on mutations.

\end{abstract}


\maketitle
\renewcommand\footnotetextcopyrightpermission[1]{} 
\pagestyle{plain}

\section{Introduction}
The success of large-scale transformer models \citep{vaswani2017attention} in Natural Language Processing ({{NLP}}), further powered by the self-supervised learning approach \citep{devlin2019bert}, has led to a paradigm shift in the way researchers and model engineers design and construct source code models. Thanks to the transformer architecture, several modern source code models such as {CodeBERT} \citep{feng2020codebert}, {GraphCodeBERT} \citep{guo2021graphcodebert}, {PLBART} \citep{ahmad2021unified}, {CodeT5} \citep{wang2021codet5}, and others have achieved state-of-the-art performance on a number of source code tasks including code completion, code summarization, and code translation. 

Recently, \citet{chen2021evaluating} introduced \codex{}, a source code model based on {GPT-3} with over 12 billion parameters.
The capabilities of the model went beyond traditional code tasks, with applications not only in code completion, code summarization, etc., but also in building simple full-scale apps, and games such as Snake and Tetris. Since, \codex{} is a descendant of the OpenAI {GPT-3} \citep{brown2020language}, it inherits much of the natural language understanding capability, which gives it the power to produce completions based on natural language prompts from the user. In fact, \codex{} is able to handle natural language prompts not only in English, but also in multiple languages such as German and Japanese, which is a commendable since users can effectively prompt the model in their own language.

The marked ingenuity of the \codex{} model is doubtlessly impressive with far-reaching applications, however, it must still be evaluated against benchmarks for the range of source code tasks it promises to handle \citep{karmakar2019establishing}. Since several of the publicly available source code repositories on GitHub may contain code that is poorly written, insecure, or downright malicious,\footnote{GitHub contains some malicious programs that can alter their environments. \citep{rokon2020sourcefinder}} large-scale source code models depending on such input data may produce code that is harmful with unanticipated consequences, both for the model developers and the end-users. Researchers found preliminary evidence that source code synthesized by \codex{} can include vulnerabilities from \texttt{MITRE}'s top 25 most dangerous software weakness list \citep{Pear21a}. Furthermore, several types of biases may exist in the training data that could lead such models to produce biased outputs. 

Therefore, rigorous testing and auditing, in multiple evaluation formats and stages, is necessary to understand the potential advantages and pitfalls of such models---in a bid to ultimately making model predictions reliable and transparent, and to open up the black boxes that are large-scale source code models.

In this paper, we evaluate the code synthesis capabilities of the \codex{} model by prompting a set of 115 Python problem statements defined on \hackerrank{}\footnote{\url{https://hackerrank.com}} and assess whether the model is able to correctly synthesize code from natural language problem descriptions. Our evaluation shows that \codex{} is indeed able to produce valid code for 96\% of the problems in a zero-shot setting, and 100\% of the problems in a few-shot setting with varying temperature values for each trial.

However, \codex{} shows signs of memorization, as several initial experiments we designed demonstrate. When the input-output specifications were deliberately omitted from the prompt, the majority of the outputs matched the specifications corresponding to the original \hackerrank{} problems. Moreover, \codex{} produces the full working code for some \hackerrank{} problems with just the first sentence of the problem statement as a prompt---even without clear task objectives. Based on these surprising observations, we propose a preliminary framework for code synthesis evaluation, and discuss their implications.

\newpage
\section{Related Work}
\label{sec:related_work}
Statistical modelling for source code has come a long way since it was first introduced, going from simple n-gram models \citep{hindle2016naturalness} to modern-day source code transformers (as surveyed by \citet{allamanis2018survey, allamanis2021surveyonline}). Transformer-based source code models, such as GraphCodeBERT \citep{guo2021graphcodebert}, PLBART \citep{ahmad2021unified}, CodeT5 \citep{wang2021codet5}, etc., have all reported state-of-the-art performance on tasks such as code completion, code summarization, code defect prediction, code clone detection, among many others. However, recent large-scale source code models with several billions of parameters, such as OpenAI's \codex{} \citep{chen2021evaluating}, and DeepMind's AlphaCode \citep{alphacode}, have begun to establish their dominance in code tasks, particularly in code synthesis. 

\textbf{Code Synthesis.} The idea of code synthesis from natural language statements is not new. \citet{ginsparg1978natural} and \citet{10.5555/31870.31881} outline and survey automatic programming systems that can carry out natural language dialogue exchange. Moreover, at the turn of the millennium, further attempts at general-purpose code generation from natural language began to surface \citep{price2000naturaljava, vadas-curran-2005-programming, nlp4nlp}. \citet{gulwani2011automating} contributes further to the field by presenting an algorithm that can synthesize short string-manipulating programs from input-output examples, and by surveying the state-of-the-art approaches \citep{gulwani2017program}.

More recently, studies by \citet{10.5555/3042817.3042840} and \citet{parisotto2016neurosymbolic} show how learning on input-output examples can be leveraged to automatically synthesize code. \citet{yin2017syntactic} improve semantic parsing and achieve state-of-the-art results. Work on code synthesis reached a new milestone when Codex was introduced by \citet{chen2021evaluating} with the promise of generating \emph{complete} snippets of code from clearly defined problem statements in natural language. While newer models such as AlphaCode \citep{alphacode} promise to take program synthesis to greater heights, \citet{austin2021program} and \citet{karmakar2021pre} have already begun exploring the limits of code synthesis and code understanding in large language models.

\textbf{Evaluation of \codex{}.} \citet{pearce2022pop} evaluate the \codex{} model to identify the purpose and capabilities of code snippets, and identify important variable names or values from code, by prompting open-ended questions to the model. The authors develop a true/false quiz framework to characterize the performance of Codex. \citet{prenner2021automatic} evaluate \codex{} on the task of automated program repair; while \citet{pearce2021openai} evaluate Codex and Jurassic J-1 models on their ability to repair insecure code in a zero-shot setting. 
\citet{Pear21a} investigated whether and to which extent code synthesized by \codex{} include vulnerable code, finding  that 40\% of the 1,692 generated snippets were vulnerable.
These studies shed light on the question answering and code repair capabilities of the \codex{} model as well as security implications, while we evaluate the model specifically on code synthesis.

\citet{tang2021solving} utilize the code synthesis capabilities of Codex to solve university-level problems in probability and statistics. 
Similarly, \citet{drori2021solving} solve algebra problems using Codex. The authors first take problems from MIT, Stanford, and Columbia University's courses, then convert them into suitable programming tasks, and then prompt the Codex model to generate code solutions. \citet{drori2022neural} go further on to solve
calculus and differential equations problems using the code synthesis capabilities of \codex{}. These studies evaluate \codex{} on code synthesis, similar to our approach, but their evaluation efforts remain limited to math problems.

\section{Experimental Considerations}
\label{sec:experimental_considerations}
In this paper, we evaluate \codex{} in a zero-shot setting on a single task of code synthesis from natural language problem statements. 
Since we evaluate \codex{} in a zero-shot setting, no further fine-tuning was done on it---in order to ascertain the raw predictive power of the model. Since \codex{} was demonstrated to be particularly proficient at code synthesis, producing legitimate code fragments from just natural language cues or prompts, we decided to evaluate it on the same code synthesis task. Furthermore, since \citet{chen2021evaluating} highlighted in their paper that \codex{} was the most competent in Python, we chose to evaluate \codex{} on the code synthesis task for the Python language.  

We received private beta access to \codex{}, which allowed us to run our evaluations. \codex{} is released in two formats: Codex-davinci
\textit{(made available as \textit{\texttt{code-davinci-001}})} 
and Codex-cushman
\textit{(made available as \textit{\texttt{code-cushman-001}})}. By default, we evaluate the Codex-davinci model, which is the larger and more capable \codex{} model, notably competent in code synthesis.

While evaluating Codex-davinci on the code synthesis task, the default settings were used: with the Temperature set to 0, the Top-P set to 1, the frequency and presence penalty set to 0, and taking only the best of 1 completions. All completions were done with an initial response length of 128 tokens, and subsequent completions were continued till no new tokens were produced. The settings remain the same for all zero-shot evaluations, with varying Temperature values used for few-shot evaluations.

To prompt the \codex{} model for synthesized code solutions, we use the problems defined on a popular competitive programming platform, \hackerrank{}---which provides a range of well-defined problem statements with conceptual explanations, examples, and input-output specifications, designed to test the Python proficiency of human programmers. We extract the natural language (English) problem statements from these problems, and if needed also the input-output specifications and examples, to formulate our prompts, and evaluate the subsequent solutions synthesized by the Codex model based on these prompts.

The prompts are presented to \codex{} as docstrings; the model automatically detects the language of choice and makes its predictions in Python. The prompts presented are straightforward with clear task objectives and input-output specifications, avoiding additional definitions, or explanations wherever possible. We retrieve the synthesized code and submit it to the \hackerrank{} platform's in-built test suite which runs its test cases to accept or reject the code solution. If the tests pass, the solution is considered correct, as it would for an ordinary human programmer.

\section{Results}
\label{sec:results}
Our evaluation shows that \codex{} is indeed capable of {resolving} a class of code synthesis problems; specifically from \hackerrank{}’s list of problems that are used to determine Python proficiency in human participants. 
The problem statements prepared for this evaluation and their corresponding code solutions generated by the \codex{} model are made available online as runnable scripts.\footnote{\url{https://github.com/giganticode/codex\_vs\_hackerrank}} Table \ref{tab:overview_hackerrank} gives an overview of evaluations carried out across prompt types and the summary of results, as well as our expectations.

\newpage
\paragraph{Full problems}
Out of 115 code synthesis problems, \codex{} correctly generates solutions to 111 of them in a zero-shot setting, and to all 115 of them in a few-shot setting \textit{($\leq$ 3)}, with a success rate of 96\% and 100\% respectively. However, there are some serious caveats. \codex{} seems to be parroting memorized code instead of actually synthesizing the solution from the problem statement. This is reflected in several situations as detailed below.

\begin{table}[t]
\caption{Overview of evaluations carried out across prompt types and summary of results.}
\label{tab:overview_hackerrank}
\begin{center}
\small
\begin{tabular}{lccccc}
\toprule
\multicolumn{1}{l}{\bf Prompt Types} & \multicolumn{1}{r}{\bf  Tested} & \multicolumn{1}{r}{\bf  Passed} & \multicolumn{1}{r}{\bf Pass\%} & \multicolumn{1}{r}{\bf Expected} \bottomspace{}\\ 
\midrule

{Full Problems}                 & 115  & 115 & 100\% & High \topspace{}\\ 
{Missing Specifications}        & 100  &  84 &  \textbf{84\%} & \textbf{Low} \topspace{}\\ 
{Missing Objectives}            &  88  &  33 &  \textbf{38\%} & \textbf{Low} \topspace{}\\ 
{Different Objectives}            &  20  &  3 &  \textbf{15\%} & \textbf{High} \topspace{}\\ 

\bottomrule
\end{tabular}
\end{center}
\end{table}

\begin{figure}[htp]
    \centering

    \begin{subfigure}{0.475\textwidth}
        \textcolor{lightgray}{\frame{\includegraphics[width=\linewidth]{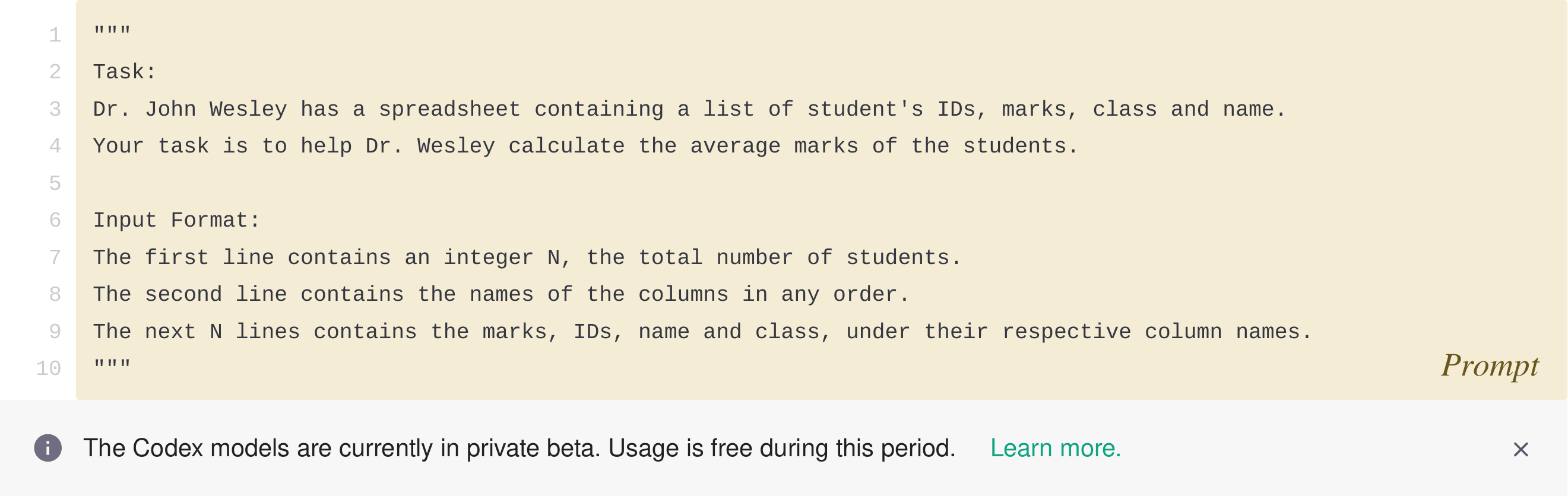}}}
        \caption{The natural language prompt (original).}
        \label{fig:missing_objectives_a}
    \end{subfigure}

    \begin{subfigure}{0.475\textwidth}
        \textcolor{lightgray}{\frame{\includegraphics[width=\linewidth]{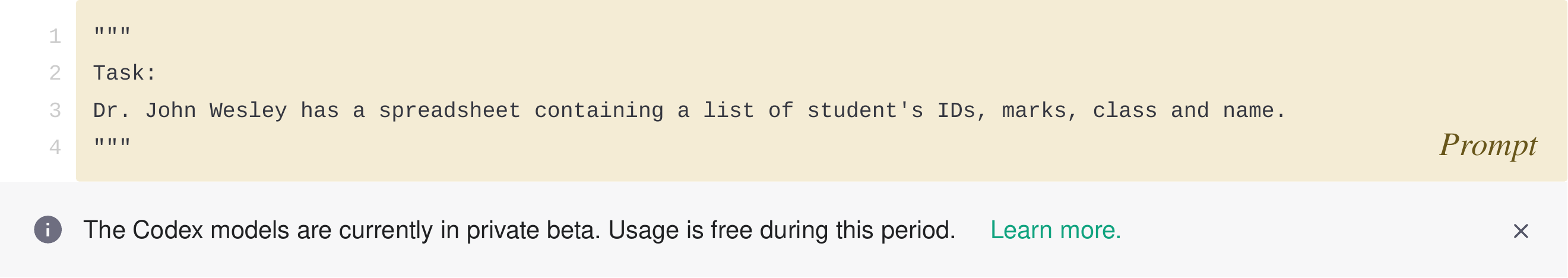}}}    
        \caption{The trimmed prompt without objectives.}
        \label{fig:missing_objectives_b}
    \end{subfigure}
    \caption{Problem statement \#57 original and trimmed.}
\end{figure}

\paragraph{Missing input/output specifications.} Input/output specifications define the format of the data provided to the problem, as well as the output expected by the test cases.  Figure~\ref{fig:missing_objectives_a} provides an example specification in lines 7 to 9. There could be many alternative input formats. This specification does not affect the logic of the problem, but still significantly affects its behaviour. Thus, a model that does not memorize  \emph{should perform badly} in the absence of specifications.

Out of 115 problem statements, nine do not have any input-output specifications, constraints, or examples included in the problem statements. For six other problems the prompt is straightforward: the model does not need input-output specifications or additional information to synthesize code even if they are explicitly provided. We exclude these 15 problems in this evaluation since the specifications either do not exist or are not strictly necessary, to better determine whether \codex{} can actually synthesize correctly-formatted and valid code even when the necessary input-output specifications are not provided. This leaves 100 problems.

For these 100 problems, we prompted \codex{} with just the problem statements \emph{without any input-output specifications}. The solution generated matched the specifications mentioned in the original problem statement in 84\% of the cases. Even for problems where the output must be structured in a specific way, 
the \codex{} model produced code snippets that matched the required output conditions as specified in the original problems on \hackerrank{}. Without the knowledge of how the input has to be read from the user and how the outputs must be structured, \codex{} seems to be generating learned code it has \textit{seen} rather than actually synthesizing it.

\paragraph{Missing objectives.} A stronger test for memorization is to remove the task objective altogether. Consider the problem statement numbered \#57, the full problem statement of which is shown in Fig. \ref{fig:missing_objectives_a}, and a trimmed version of the same problem statement is shown in Fig. \ref{fig:missing_objectives_b} that has no task objective. Surprisingly, \codex{} predicts a valid code snippet corresponding to the full problem statement as defined on the \hackerrank{} platform with matching input specifications, matching task objective, and passing all the test cases, even with a trimmed prompt consisting only of: \textit{\enquote{Dr. John Wesley has a spreadsheet containing a list of student IDs, marks, class, and name}}. A model unaware of the objective (calculating the average mark) \emph{should fail}; \codex{} is clearly memorizing.

Out of 115 problem statements, 27 have some sort of program objective mentioned in the very first sentence, such as, %
problem \#41 ``\textit{You are given a complex z, convert it into polar coordinates}''; or problem \#96 ``\textit{You are given a valid xml, and you have to print the maximum level of nesting}''. For most of these 27 problem statements additional input-output specifications, or explanations are required to actually solve the problem. But, if one gives \codex{} the benefit of the doubt, one would assume that \codex{} can synthesize code in a fair manner just from the program objective mentioned in the very first sentence, so we excluded these problems from this evaluation. %

Out of the remaining 88 problems where no program objective is present in the first sentence, we found 33 instances where Codex produces the full and valid \hackerrank{} solutions to the problems just from the first sentence, including matching input-output specifications, stub code, and comments ($\sim$40\% of the problems).

\paragraph{Different Objectives.} We also assessed the impact of memorization on generalization. 
For example, problem statement \#32 asks for the \textit{sum of the elements in set A} to be printed as output. Upon prompting the \codex{} model with problem statement \#32, it predicts the correct solution as expected.  %
However, when we intentionally modify the problem statement to another variant, where the \textit{product of the elements in set A} is to be printed, the model fails: it still presents the same code solution from the original problem statement (the sum) when prompted for the variant (the product). This suggests that the output was likely learned from several appearances of the stated problem in the training data: prompting a modified variant triggers the model to output the memorized code. 

We found 20 problems in our original set for which defining such variants was straightforward. We were surprised to see that \codex{} fails to predict the correct code solutions for 17 out of 20 variants of problem statements, often producing code solutions corresponding to the original unmodified \hackerrank{} problems. 
This suggests that memorization is strong enough to impede generalization.

\resultbox{Codex shows clear signs of memorization: it often generates complete and correct code even when significant information is missing in the prompts. This impacts generalization: \codex{} often produces erroneous code when faced with similar, but distinctly different problems to the ones seen in training.}

\newpage
\section{Towards a Novel and Systematic Evaluation Benchmark}

Our results suggest that Codex might fail to generalize, especially when well-known problem statements (i.e., likely seen during training) are changed in subtle ways. To truly evaluate a model's capability of general code synthesis, we propose a novel type of evaluation benchmark, where the problem statements contain mutation points that can take on different values (Figure~\ref{fig:grid1}). 

Instead of a single problem statement, we generate and evaluate the space of all possible value combinations that can be passed into the mutation points. Depending on the number of mutation points and values, a single problem statement can thus have hundreds of variants. Since writing test cases for each variant is not feasible, we use an \emph{oracle}---a correct reference implementation, to assess the correctness for all variants by comparing a number of predefined inputs the outputs of the synthesized solution and the oracle.

By calculating marginal success rates for each value at a mutation point, it is possible to find values that frequently cause incorrect output. Figure~\ref{fig:grid1} shows an example problem with two variable mutation points, (\texttt{double\_class}, \texttt{triple\_class}) that can take several values, e.g., '\texttt{letter}', '\texttt{number}', or '\texttt{question mark}'. For every mutation point and value combination, we generate a new problem variant. We have so far prepared and evaluated over ten such problems---with up to eight mutation points in each and up to 250 variants for a single problem. Several of these problems have been adapted from HumanEval~\citep{chen2021evaluating}, the evaluation set written by OpenAI to evaluate the Codex model. HumanEval should be less susceptible to memorization issues since OpenAI ensured it was not part of the training data. We briefly discuss three examples. 

\textit{Binary operators.} One of the HumanEval problems (problem 11) requires the synthesis of the binary XOR operator. Our version of this problem tests for several different binary operators, including OR, NOR, AND, NAND, XNOR/XAND. While Codex is able to synthesize a correct solution for the original problem with XOR, it fails for NAND, XNOR and XAND.

\textit{Fizz Buzz.} We adapted the well-known Fizz Buzz problem, a simple programming exercise often used in programming interviews, where a range of numbers need to be tested for divisibility by two pre-defined factors (usually three and five). We control the factors through template variables and test for a range of different factors, including writing numbers as numerical digits versus English words. Codex is able to correctly synthesize a solution for almost all combinations, but shows signs of brittleness. In some settings it interprets \texttt{twenty-three} correctly as \texttt{23}, while in others it interprets it as a subtraction (2 minus 3), leading to invalid code. 

\textit{String encoding.} For a problem involving encoding a string as a hexadecimal hash-string using a hash algorithm controlled by different mutation values (e.g., SHA1, SHA512, MD5) we find that Codex is able to synthesize correct solutions for 67\% of the prompted variants, and producing erroneous code for the rest.

\resultbox{Systematic evaluations of problems show that beyond memorization issues, the output of Codex varies significantly when problem descriptions are altered, implying that further investigation is needed on the code synthesis capabilities of Codex.}

\begin{figure}[t!]
    \begin{lstlisting}[language=Python, numbers=none, belowskip=-1.0 \baselineskip, frame=single]
  def double_letters(string):
    """ Given a string, return the string, doubling every {(*\textbf{\color{gridvar}double\_class}*)} character, and tripling every {(*\textbf{\color{gridvar}triple\_class}*)} character. All other characters should be output a single time each. """
    \end{lstlisting}
    \caption{A problem with two \textit{variable} mutation points}
    \label{fig:grid1}
\end{figure}

\section{Conclusion} %
\label{sec:implications}
Our preliminary study shows that \codex{} is subject to memorization and generalization issues, even in some relatively simple settings. This clearly calls for additional investigation of the problem, since the use and adoption of \codex{} and other large source code language models as foundation models %
can have a far-reaching impact on the way future code is produced, maintained, and used. %
We highlight some of the issues and risks associated with source code models; a broader discussion of risks for NLP language models, including social biases, was done by
\citet{stochastic2021parrots}. \\[-10pt] %

\textit{Issues with scientific evaluations.} 
Previous machine learning models were already sensitive to code duplication \cite{allamanis2019adverse}. \codex{} has been trained on 55 million GitHub repositories \cite{chen2021evaluating}; moreover, which repositories it was trained on is not publicly known. Many existing code datasets have been gathered from open-source code repositories available on platforms like GitHub, it is likely that evaluating large-scale code models on these datasets can result in inaccurate or biased evaluation outcomes, especially since Codex can memorize code very well. For instance, prompting \codex{} to fix a buggy snippet of code can result in correct code \cite{prenner2021automatic}; this indicates that either \codex{} is capable of program repair, or that it has seen the correct code snippet, possibly multiple times, during training. %
Therefore, going forward, researchers must be careful in evaluating large-scale source code models with datasets derived from GitHub to avoid biased evaluation results, especially datasets containing commonly-used or well-known code snippets or algorithm implementations. \\[-10pt] %

\textit{Issues with memorization for source code language model users.}
In some cases, \codex{} appears to be working more like a code retrieval engine rather than doing source code synthesis. This can cause issues, especially when the model outputs a solution that is close to, \emph{but not exactly} what was needed, as happened with several modified \hackerrank{} problems. Users of the model must carefully review the output to ensure that the generated code is indeed what they expect. While this advice sounds obvious, and the version of Codex integrated in Visual Studio Code is clearly labelled as a "Copilot", we think that the issues that \codex{} exhibited in our preliminary study show that the problem should not be underestimated. Additional research is needed to make source code synthesis more reliable, such as the Synchromesh framework \cite{poesia2022synchromesh}.

Further issues related to memorization concern privacy. Source code on GitHub may contain sensitive information such as API tokens, secret keys, or even passwords \citep{7180102}. Although the onus is upon the users to keep their information \textit{private}, 
models such as \codex{} may deepen the concern due to their propensity to memorize. \\[-10pt]

\textit{Future work.} We presented a simple but effective framework to systematically evaluate the performance of large language models such as \codex{}. We plan to significantly expand the number of problems, to conduct a thorough analysis of \codex{} and other code synthesis models, and to develop it as a novel evaluation benchmark.  

\bibliographystyle{ACM-Reference-Format}
\bibliography{sample-base}

\end{document}